# Anisotropic strain variations during the confined growth of Au nanowires

Giuseppe Abbondanza[1,2], Andrea Grespi[1,2], Alfred Larsson[1,2], Lorena Glatthaar[3], Tim Weber[3], Malte Blankenburg[4], Zoltan Hegedüs[4], Ulrich Lienert[4], Herbert Over[3], Edvin Lundgren[1]

1 Division of Synchrotron Radiation Research, Lund University, Professorsgatan 1, 22100 Lund, Sweden

2 NanoLund, Lund University, Professorsgatan 1, 22100 Lund, Sweden

3 Institute of Physical Chemistry, Justus Liebig University, Heinrich-Buff-Ring 17, 35392 Giessen, Germany

4 Deutsches Elektronen-Synchrotron DESY, Notkestr. 85, 22607 Hamburg, Germany

**Abstract**

The electrochemical growth of Au nanowires in a template of nano-porous anodic aluminum oxide was investigated *in situ* by means of grazing-incidence transmission small- and wide-angle x-ray scattering (GTSAXS and GTWAXS), x-ray fluorescence (XRF) and 2-dimensional surface optical reflectance (2D-SOR). The XRF and the overall intensity of the GTWAXS patterns as a function of time were used to monitor the progress of the electrodeposition. Furthermore, we extracted powder diffraction patterns in the direction of growth and in the direction of confinement to follow the evolution of the direction-dependent strain. Quite rapidly after the beginning of the electrodeposition, the strain became tensile in the vertical direction and compressive in the horizontal direction, which showed that the lattice deformation of the nanostructures can be artificially varied by an appropriate choice of the deposition time. By alternating sequences of electrodeposition to sequences of rest, we observed fluctuations of the lattice parameter in the direction of growth, attributed to stress caused by electromigration.. Furthermore, the porous domain size calculated from the GTSAXS patterns was used to monitor how homogeneously the pores were filled.

The electrochemical deposition of metal in hard templates of nano-porous anodic aluminium oxide (NP-AAO) represents a facile and versatile bottom-up method for the fabrication of ordered arrays of metal nanowires[1–7]. NP-AAO is a material synthesised in the anodization of Al and, through its self-assembling honeycomb structure, large areas of material can be fabricated by simply exposing Al to the electrolyte solution under anodizing conditions, making the template method an easily scalable procedure for industrial purposes[8,9]. In previous research, we presented a facile template-assisted route for the fabrication of Au nanowires in NP-AAO and we characterised them by *ex situ* electron microscopy and x-ray diffraction[7].

The *hard* template method (as opposed to the *soft* template method, which relies on the assistance of colloidal aggregates[10]) consists in electrodepositing metal in an inorganic nano-porous medium. Since the electrodeposited metal adopts the shape of the pores and the length is proportional to the deposition duration, it is possible to fabricate nanomaterials with controlled morphology and aspect ratio. Furthermore, the electrodeposited nanostructures can be released by selectively dissolving the NP-AAO template to obtain nanostructures dispersed in solution or ordered arrays of up-standing nanowire forests on a substrate. The release from NP-AAO, by the action of selective etchant like NaOH, exposes the large surface area per unit volume which gives nanomaterials their extraordinary properties. Another advantage of the hard template-method is that the surface of the electrodeposited metal is surfactant-free, which means that de-capping stages are not necessary for applications where the availability of surface sites is relevant, such as in catalysis[11–13]. Au nanowires in NP-AAO might also be employed as nanoelectrode arrays which enhance the mass transport and lead to steady-state voltammograms with sigmoidal shape, unaffected by the diffusion limit[14].

It has been previously demonstrated that Au nanowires deposited in NP-AAO templates have an anisotropic size-dependent strain distribution[7]: the lattice parameter is larger in the direction of growth than in the direction of confinement and the strain is inversely proportional to the pore radius, which suggests that the interatomic distances of the electrodeposited nanomaterials can be fine-tuned by selecting an appropriate template pore radius. Similar findings have been previously reported about Pd[6] and Sn[15,16] nanowires. The potential applications of a tuneable size-dependent strain range from catalysis, where it has been shown that small variations of the interatomic distances can alter the catalytic activity of materials[17–22], hydrogen solubility in the crystal lattice of solids[23] to strain-engineering of functional nanowires for devices[24].

The anisotropic strain effect has been attributed to the sum of two contributions: surface stress and growth stress[16]. However, the origin of the growth stress and its interplay with the electrochemical conditions are not fully understood. For instance, *in situ* stress measurements of electrochemically grown thin films revealed stress relaxation caused by the interruption of growth[25], which suggested that the operating electrochemical conditions have an influence on the strain state of the deposited material. To our knowledge, there is a lack of similar studies applied to the template-assisted growth of nanowires in NP-AAO. For this reason, we investigated the electrochemical growth of Au nanowires by means of GTWAXS, to learn about the structure-function relation regarding the interatomic distances in the direction of growth and in the direction of confinement, and simultaneously we monitored the progress of the growth by means of GTSAXS, XRF and 2D-SOR, which is a technique that has been proved useful in bridging macroscopic and atomic-scale observations[26–29]. High-energy synchrotron x-rays were necessary to study the encapsulated growth of the Au nanowires with a time resolution compatible with the growth rate.

A detailed description of the template preparation, the electrodeposition method, and the Au nanowires synthesis protocol is available in previous research[7]. In brief, the NP-AAO template was fabricated prior to the synchrotron experiment by anodizing a top-hat shaped Al sample (99.999%, Surface Preparation Laboratory, Netherlands) with a diameter of 6 mm, using an electrochemical holder to expose only the top surface of the specimen. The insulating barrier layer at the pore bottom was thinned by decreasing the anodizing potential[3]. The electrochemical growth was conducted using the pulse electrodeposition method in an electrolyte solution containing $HAuCl_4$ in a phosphate buffer with neutral pH, pumped through an electrochemical flow-cell specifically designed for *in situ* x-ray measurements[28]. We alternated sequences of electrodeposition lasting 2 min with sequences of rest lasting 3 min, where the electrochemical cell was left at the open-circuit potential, aiming to study any strain variation caused by the absence of an external electric field. During the electrodeposition sequences, we measured the potential across a shunt resistor with a resistance of 25 Ω, connected in series with the electrochemical cell. The measured potential was used to calculate the current flowing through the cell using Ohm's law.

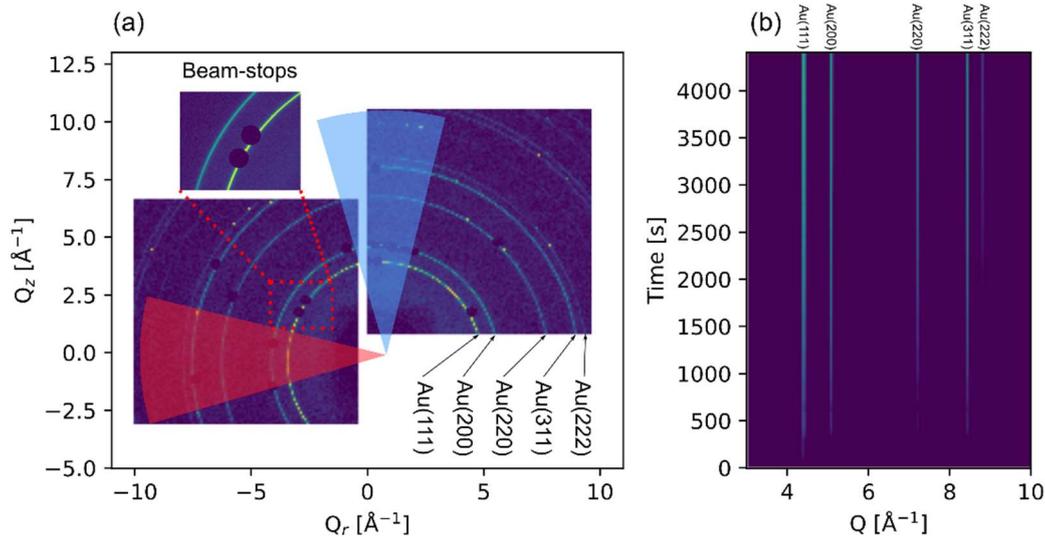

**Figure 1.** (a)GTWAXS detector images collected at the end of the electrodeposition, (b) waterfall plot of the GTWAXS patterns measured *in situ*.

The synchrotron measurements were conducted at the Swedish Material Science beamline P21.2 (Petra III, DESY), using an x-ray beam with an energy of 68.5 keV and a size of 13 μm × 100 μm (vertical × horizontal). The experiment was performed in grazing-incidence transmission geometry[30], using an incidence angle of 0.15°. For a detailed description of the experimental setup, its calibration and the data processing procedure, we refer the reader to a recent manuscript about the electrochemical growth of Pd nanowires, where we employed the same setup and a similar template[31].

A representative GTWAXS pattern collected at the end of the electrodeposition is shown in Fig. 1 (a). The Bragg reflections originating from the polycrystalline Al substrate were screened using tungsten beam-stops and rectangular lead foils. The less intense Al reflections which did not need screening were masked out from the azimuthal integration during the data processing. The intensity distribution along the Au diffraction rings in Fig. 1 (a) is homogeneous throughout the course of the deposition, which reflects the isotropic random orientation of the Au crystallites. Fig. 1 (b) is a waterfall plot where every horizontal line is a powder diffraction pattern obtained by integrating the GTWAXS images over the whole azimuthal range available, using the "Multi-geometry" module in pyFAI[32,33]. Fig. 1 (b) gives an overview on the progressive appearance of the Au face-centered cubic (fcc) phase during the course of the electrodeposition.

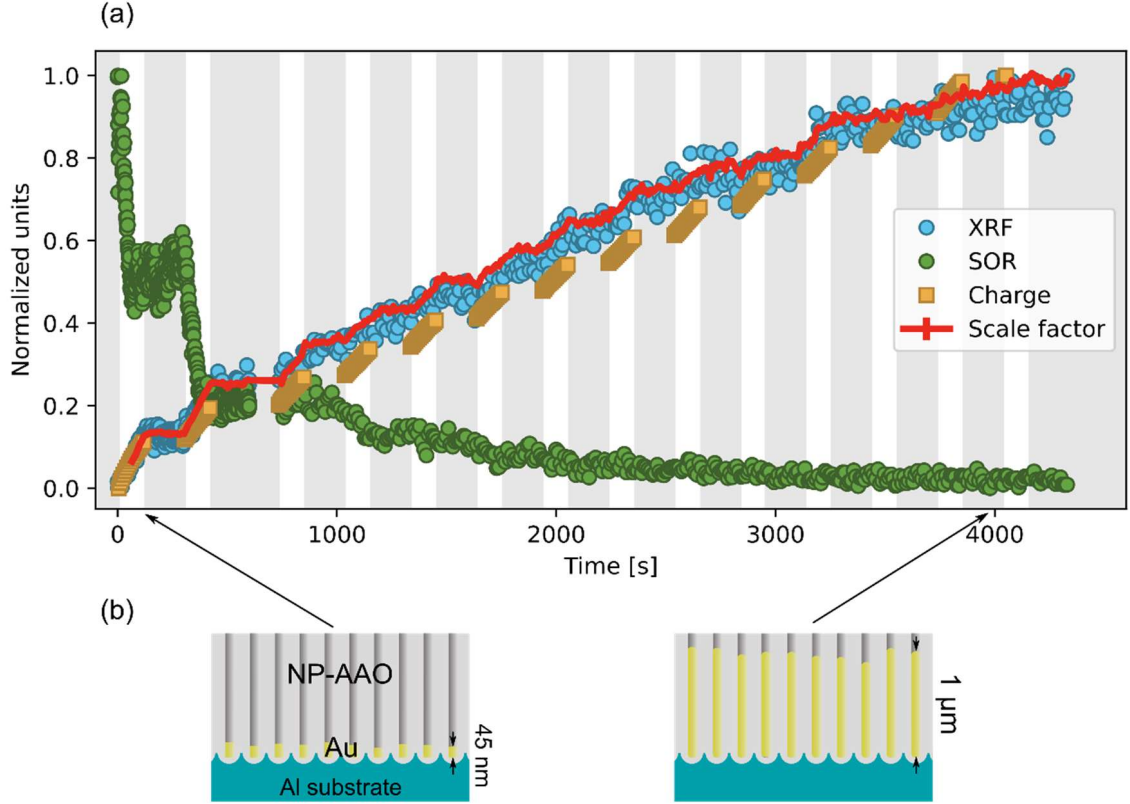

**Figure 2.** (a) XRF, SOR, cathodic charge and GTWAXS scale factor as a function of time, during the electrodeposition of Au in NP-AAO. The electrodeposition is ongoing in the white areas and interrupted in the grey areas. (b) Schematic of the nanostructures in NP-AAO at the beginning and the end of the deposition. The length of the nanowires at the instants in time indicated by the arrows are based on length estimations available in previously published *ex situ* electron microscopy observations[7].

To study the anisotropy of the system, we integrated the GTWAXS patterns over azimuthal slices oriented along the vertical and the horizontal axis, shown as the blue and red shaded areas in Fig. 1 (a), respectively, to obtain one-dimensional powder diffraction patterns. The width of the azimuthal slices was 15°. We found that such azimuthal range was wide enough to provide a good signal-to-noise ratio and sufficiently thin to study anisotropy effects. In the patterns obtained by integrating in the horizontal direction, the scattering vector always lies in the horizontal plane, i.e., it always lies in the direction of confinement. On the other hand, in the patterns obtained by integrating over the vertical direction, the angle between the direction of growth and the scattering vector is never zero but it is sufficiently small. For instance, the Au(111) reflection appears at a Bragg angle θ of 2.20° (with an x-ray beam energy of 68.5 keV). The growth proceeds in the direction of the pores, which are vertically aligned with the specimen surface normal. This means that the angle between the growth direction and the scattering vector is θ-α=2.05°, where α is the incidence angle (which was 0.15° in this setup). This angle is sufficiently small to allow an investigation of the interatomic distances in the direction of growth with good approximation.

Fig. 2 (a) is a plot of the XRF signal, arising from the Au Lα emission line, the 2D-SOR signal integrated over the whole reflecting sample surface, the total cathodic charge that has passed through the working electrode and the scale factor of the Au fcc phase, obtained by the sequential Rietveld refinement (performed using GSAS-II[34]) of the powder diffraction patterns in the vertical direction. The white and grey regions of the plot represent the periods of time while the deposition was ongoing and interrupted, respectively. While the scale factor is proportional to the amount of crystalline material deposited, the XRF intensity is sensitive to both crystalline and amorphous phases. The fact that the XRF and the scale factor increase similarly suggests that there is no amorphous Au forming during the growth and that only crystalline Au is deposited.

To illustrate the nanowire size evolution over time, Fig. 2 (b) depicts the deposited Au nanostructures at two instants in time: close to the beginning and close to the end of the deposition. The size estimations are based on previous *ex situ* observations by scanning electron microscopy[7]. The linear increase of the XRF and the scale factor suggest that the growth rate is constant through the whole electrodeposition.

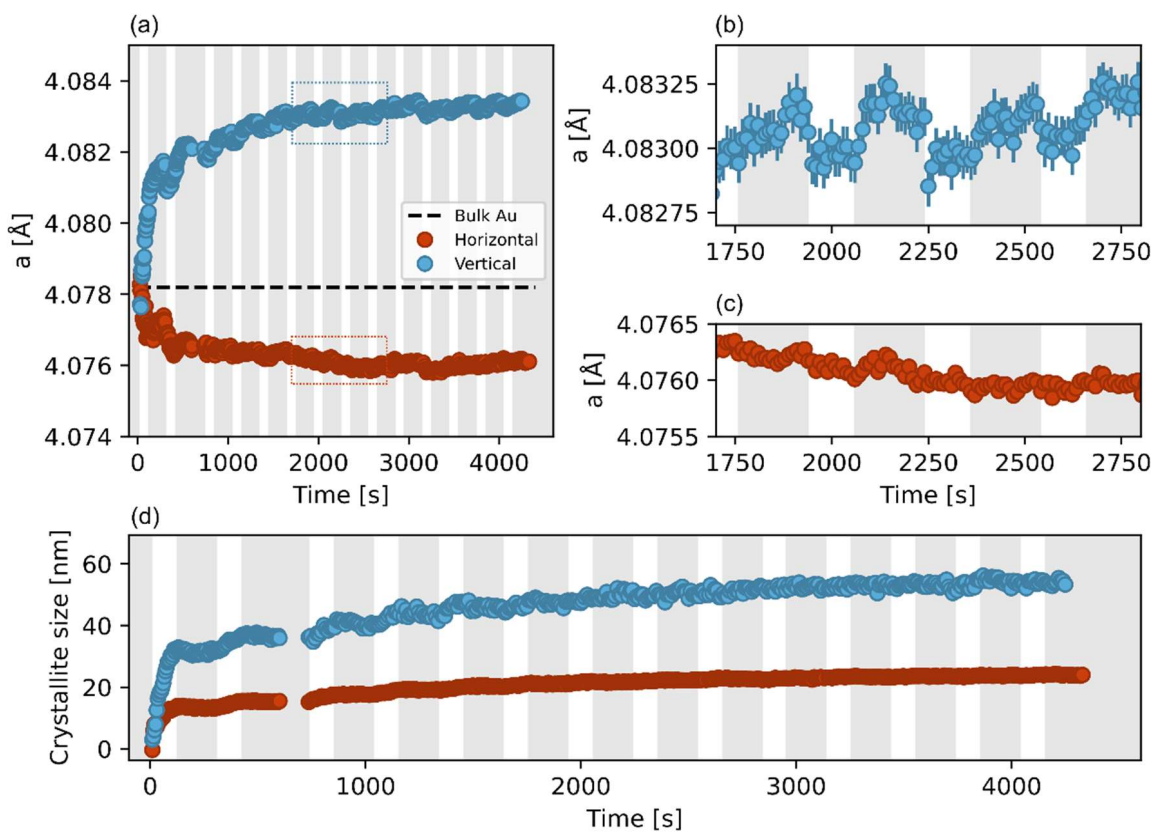

**Figure 3.** (a) Au lattice constant as a function of time in the horizontal and vertical directions, (b) magnified view of the blue box in (a), (c) magnified view of the red box in (a), (d) crystallite size as a function of time, in the horizontal and the vertical directions.

The cathodic charge as a function of time in Fig. 2 (a) was obtained by integrating the total negative current flowing through the working electrode from the beginning of the deposition to any instant in time. The cathodic charge is also a measure of the total amount of material in the pores as a function of time and the fact that it scales with the XRF and GTWAXS is an indication that the rate of side reactions, occurring at the same time with the electrodeposition, is constant as a function of time.

The 2D-SOR signal shows a rapid decrease in the first 600 s of the deposition, followed by a more gradual decrease until the end of the deposition. As the refractive index of Au (in the optical wavelength regime) is higher than that of alumina, one would expect the reflectance to increase, as predicted by the Fresnel laws for the case of unpolarised light with normal incidence to a reflecting surface. The decrease of reflectance shown in Fig. 2 (a), however, evidences that reflection is not the only light-matter phenomenon involved. We surmise that the reflectance decrease is caused by light absorption by plasmon resonance and by light scattering. To support this hypothesis, UV/Vis absorbance spectra of Au nanowires in NP-AAO, found in the scientific literature, show variations as a function of the aspect ratio[35]. It has also been shown that scattering events participate to the extinction of the incident light[36]. However, our 2D-SOR setup uses a monochromatic red LED as a source and the measurement of absorbance spectra is beyond the purpose of the technique.

From the powder diffraction patterns integrated in the horizontal and vertical direction, we estimated the lattice parameter by means of sequential Rietveld refinement, shown in Fig. 3 (a). Quite rapidly from the beginning of the electrodeposition, the lattice constant in the vertical direction increases and becomes higher than the lattice constant of bulk Au. Conversely, the lattice constant in the horizontal direction decreases and becomes smaller than bulk Au. This anisotropic effect is analogous to what was reported in literature about the size-dependent strain in electrochemically fabricated Sn nanowires in NP-AAO[5,15,16]. In a similar way to the case of Sn nanowires, the anisotropic strain can be explained by a combination of surface stress and growth stress, where the latter was attributed to the presence of defects but it can also be due to deviations from the natural pathways of growth due to confined environment[37,38]. As the deposition proceeds, the anisotropic effect becomes more pronounced, which suggests that the strain state depends not only on the pore radius but also on the nanowire length.

Fig. 3 (b) shows a magnified view of the blue highlighted area in Fig. 3 (a). Here, small fluctuations of the lattice constant were observed, which have the same periodicity as the on/off state of the electrodeposition, represented by the white and grey areas of the plot. Although the origin of the stress that causes this deformation is unclear, a possible explanation is the electromigration of Au, which is the thermally-assisted mass transport of ions under the influence of an electric field. This phenomenon is explained by the unbalance of electrostatic and electron-wind forces exerted on metal ions[39]. This mechanism has been proved to cause stress gradients and mechanical failure in Au nanowires[40]. *In situ* x-ray diffraction studies on bulk Cu[41], Sn[42] and ferroelectric thin films[43] revealed similar variations of the lattice parameter driven by the presence of an electric field. Fig. 3 (c) is amagnified view of the lattice constant in the horizontal direction, in the same time window as in Fig. 3 (b). Here, the lattice parameter fluctuates less than in the vertical direction. The fact that the strain fluctuations are more pronounced along the direction of growth, which is parallel to the electric field lines strengthens the electromigration hypothesis[44].

The crystallite size, extracted from the powder diffraction patterns in the vertical and horizontal direction, is reported in Fig. 3 (d). At any point in time, the crystallite size is larger in the direction of growth than in the direction of confinement, which is due likely to a constrain in the horizontal plane by the confinement of the pores. In fact, the crystallite size in the horizontal plane never exceeds 25 nm, which is the pore diameter of the template anodized in sulphuric acid. The rapid increase of the crystallite size during the first sequence (120 s) of deposition can be attributed to coalescence phenomena of the grains nucleating in the pores bottom[45,46]. The slower kinetics of grain growth at longer deposition times can be explained by the "generalized parabolic grain growth model", which describes the curvature of the grain boundaries as the driving force of the grain expansion[47]. The grain growth rate predicted by this model diminishes to zero for $t \rightarrow \infty$, as the curvature radii of the grains decrease over time[48].

GTSAXS provided useful complementary information on the progress of the electrodeposition. The peaks of the GTSAXS patterns collected were each fitted by a Voigt profile and the background was fitted by a third-order Chebyshev polynomial. The sum of the peak amplitudes was used as a scale factor of the overall intensity of the GTSAXS patterns, while the peak widths were used to determine the porous domain size, using the Scherrer's equation. Fig. 4 is a plot of the porous domain size and the scale factor as a function of time. The domain size at the time zero is a measure of the porous order of the empty template. While the scale factor increases monotonically as a function of time, the domain size decreases to a local minimum after approximately 240 s from the beginning of the electrodeposition. This decrease can be attributed to the inhomogeneous pore filling by the electrodeposited Au. This observation is coherent with previous research about the electrodeposition of Sn[4]. The reason of the inhomogeneous filling is attributed to the way the template was prepared: in the early stages of deposition in templates treated by barrier layer thinning (such as the one used in this work), a heterogeneous nucleation of nanocrystals occurs at the pore bottoms, preferentially in those with a thinner barrier layer[2].

However, there is an increase of the domain size after 240 s from the beginning of the electrodeposition. This can be explained by the fact that, as the pores are being filled and the scale factor increases, the GTSAXS patterns are dominated by scattering amplitude originating from the Au nanowires rather than from the empty pores. Ideally, if the Au homogeneously filled the pores to the top, the domain size would increase back to the original value of 235 nm. Instead, it plateaus around 170 nm, which is a consequence of the heterogeneous pore filling. This finding is supported by previous *ex situ* electron microscopy observations of heterogeneous nanowire height, where, at the end of the deposition, the mean nanowire height was 1049 ± 205 nm (relative standard deviation of 20%).

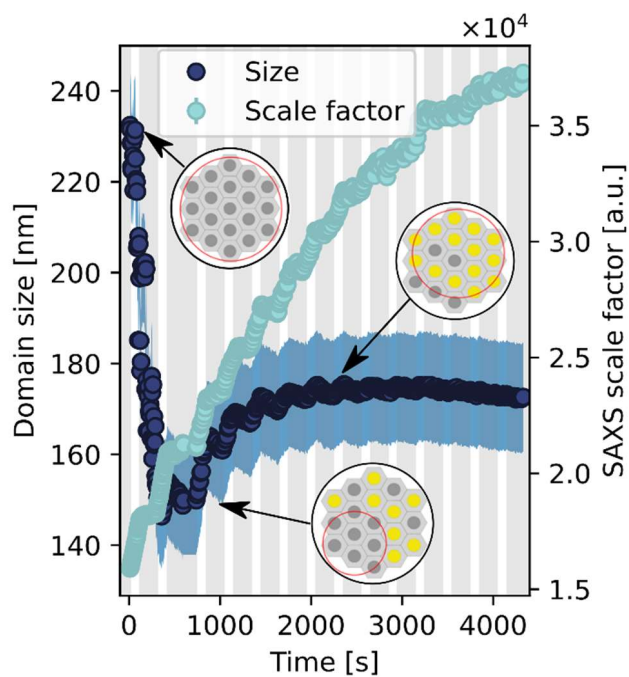

**Figure 4.** Domain size and intensity scale factor extracted from GTSAXS patterns. The insets are schematic drawings of the top view of the NP-AAO template as it is filled with Au. The highlighted areas represent the scattering domain.

In conclusion, we have followed the electrodeposition of Au into NP-AAO *in situ* using GTWAXS, GTSAXS, XRF and 2D-SOR and gained time-resolved information on the structural progress of the deposition. The lattice parameter anisotropy evolves rapidly after the start of the electrodeposition and subtle variations in the direction of growth were observed, possibly linked to the action of the electric field during growth.

These findings suggest that it is possible to artificially select the strain state of the Au nanowires by an appropriate choice of deposition duration, which might be particularly useful in the strain-engineering of nanoparticles for catalysis. For instance, it has recently been found that Au nanoparticles under compressive strain exhibit an enhanced selectivity towards electrochemical $CO_2$ reduction[49].

In future studies, we could selectively dissolve the template and at the same time monitor the lattice parameter of Au *in situ*, to observe any strain relaxation induced by the release from the template and determine if the deformations induced by growth under confinement are plastic or elastic. In addition, we could investigate the Au nanowires growth by combining our electrochemical flow-cell with fiber optics-based *in situ* UV/Vis specular reflection spectroscopy[50] to validate the hypothesis of plasmon absorption introduced in this work.

**Acknowledgements**

We acknowledge DESY (Hamburg, Germany), a member of the Helmholtz Association HGF, for the provision of experimental facilities. Parts of this research were carried out at PETRA III at the Swedish Materials Science beamline P21.2. Beamtime was allocated for proposal I-20211146 EC. We would like to thank Sven Gutschmidt for assistance in setting up the experiment.

This work was financially supported by the Swedish Research Council through the Röntgen-Ångström-Cluster 'In-situ High Energy X-ray Diffraction from Electrochemical Interfaces (HEXCHEM)' (Project No. 2015-06092), by the 'Atomic Resolution Cluster'- a Research Infrastructure Fellow program of the Swedish Foundation for Strategic Research, and project grant 'Understanding and Functionalization of Nano Porous Anodic Oxides' (Project No. 2018-03434) by the Swedish research council. We acknowledge financial support by NanoLund.